\documentclass[aps,prl,showpacs]{revtex4}
\usepackage{graphicx}
\usepackage{wasysym}

\begin{document}
\draft
\def\be{\begin{equation}}
\def\ee{\end{equation}}
\def\bfi{\begin{figure}}
\def\efi{\end{figure}}
\def\bea{\begin{eqnarray}}
\def\eea{\end{eqnarray}}
\def\cen{\centering}

\title{Non trivial behavior of the linear response function in phase ordering kinetics\footnote{Talk delivered by Marco Zannetti}}

\author{Federico Corberi\footnote{corberi@na.infn.it},
Nicola Fusco\footnote{nicola.fusco@sa.infn.it}, Eugenio
Lippiello\footnote{lippiello@sa.infn.it} and Marco
Zannetti\footnote{zannetti@na.infn.it}}

\address{Istituto Nazionale per la Fisica della Materia, Unit\`a di
Salerno and Dipartimento di Fisica ``E. R. Caianiello'',
Universit\`a di Salerno, 84081 Baronissi (Salerno), Italy}

\begin{abstract}
Drawing from exact, approximate and numerical results an overview
of the properties of the out of equilibrium response function in
phase ordering kinetics is presented. Focusing on the zero field
cooled magnetization, emphasis is on those features of this
quantity which display non trivial behavior when relaxation
proceeds by coarsening. Prominent among these is the
dimensionality dependence of the scaling exponent $a_{\chi}$ which
leads to failure of the connection between static and dynamic
properties at the lower dimensionality $d_L$, where $a_{\chi}=0$.
We also analyse the mean spherical model as an explicit example of
a stochastic unstable system, for which the connection between
statics and dynamics fails at all dimensionalities.
\end{abstract}
\maketitle

\section{Introduction}

It is a great pleasure to present this talk on the occasion of the
60th birthday of Francesco Guerra. My interest in relaxation
phenomena, in fact, initiated quite a few years ago in Salerno
where Nelson stochastic mechanics was a subject very much
cultivated under the guide of Francesco. I have benefited a lot
from his teaching and from the very stimulating atmosphere he was
able to create about the applications of stochastic methods to
physics in general.

Coming to the subject of this talk, phase ordering~\cite{Bray94}
is usually regarded as the simplest instance of slow relaxation,
quite useful for a first and easy to understand approach to
concepts like scaling and aging, which are the hallmarks of glassy
behavior~\cite{Cugliandolo2002}. However, next to the similarities
there are also fundamental differences which require to keep phase
ordering well distinct from the out of equilibrium behavior in
glassy systems, both disordered and non disordered. The main
source of the differences is the simplicity of the free energy
landscape in the case of phase ordering compared to the complexity
underlying glassy behavior. This leads to a simple
low temperature state for phase ordering systems as
opposed to the complexity of replica symmetry breaking for glasses
(at least in the mean field picture of spin glasses). An analogous
sharp distinction is believed to exist also in the behavior of the
out of equilibrium linear response function. As a matter of fact,
in the framework where static and dynamic properties are
connected~\cite{Franz98}, systems may be classified on the basis
of the fluctuation dissipation relation~\cite{Ricci99}.

In the jargon of slow relaxation, phase ordering is frequently
referred to as coarsening. So, if coarsening is associated to
simplicity and triviality of behavior what is there to investigate
about it? Besides the obvious motivation that the basic,
paradigmatic cases need to be thouroughly understood, an
additional reason, among others, is that in some cases the
existence of complex slow relaxation is identified through the
exclusion of coarsening. This requires precise and reliable
knowledge of all what goes on when relaxation proceeds by
coarsening. An example comes from the long standing controversy
about the nature of the low temperature phase of finite
dimensional spin glasses. One recent argument in favour of replica
symmetry breaking is that the observed behavior of the response
function is incompatible with coarsening~\cite{Franz98,Ricci99}.
This might well be the case. However, for the argument to be
sound, the understanding of the out of equilibrium behavior of the
response function during phase ordering needs to be up to the
level that such a delicate issue demands.

It is the purpose of this talk to present an overview of the
accurate investigation of the response function in phase ordering
that we have carried out in the last few years. Focusing on the
zero field cooled magnetization (ZFC), which is one instance of
integrated response function, it will be argued that the response
function in phase ordering systems is not as trivial as it is
believed to be and, after all, it is not the quantity best suited
to highlight the differences between systems with and without
replica symmetry breaking. In fact, as we shall see, phase
ordering and therefore a replica symmetric low temperature state
are compatible with a non trivial ZFC. When this happens there is
no connection between static and dynamic properties. Phase
ordering systems offer examples of two distinct mechanism for the
lack of this important feature of slow relaxing systems, the
vanishing of the scaling exponent of ZFC and stochastic
instability.

\section{Phase ordering}

Let us first briefly recall the main features of a phase ordering
process. Consider a system, like a ferromagnet, with order
parameter (vector or scalar, continous or discrete) $\phi(\vec x)$
and Hamiltonian ${\cal H}[\phi(\vec x)]$ such that below the
critical temperature $T_C$ the structure of the equilibrium state
is simple. For example, in the scalar case, there are two pure
ordered states connected by inversion symmetry. The form of the
Hamiltonian can be taken the simplest compatible with such a
structure, like Ginzburg--Landau--Wilson (GLW) for continous spins
or the nearest neighbors Ising Hamiltonian for discrete spins.

In the following we will be interested in the space and time
dependent correlation function

\be
   C(\vec{r},t,s)=\langle\phi(\vec x,t)\phi(\vec x',s)\rangle-
\langle\phi(\vec x,t)\rangle\langle\phi(\vec
x',s)\rangle\label{1.1}
\ee

where the average is taken over initial condition and thermal
noise, $\vec{r}= \vec{x} - \vec{x'}$ and $t \geq s \geq 0$ are two
times after the quench. The conjugated linear response function is
given by

\be
   R(\vec{r},t,s)=\left.\frac{\delta\langle\phi(\vec{x},t)\rangle}{\delta h(\vec{x'},s)}\right|_{h=0}\label{1.2}
\ee

and ZFC is defined by

\be
   \chi(\vec{r},t,t_w)= \int_{t_w}^t ds R(\vec{r},t,s).\label{1.3}
\ee

\subsection{Dynamics over phase space: equilibration versus
falling out of equilibrium}

For a temperature $T$ below $T_C$, in the thermodynamic limit, the
phase space $\Omega=\{[\phi(\vec x)]\}$ may be regarded as the
union of three ergodic components~\cite{Palmer} $\Omega=\Omega_+
\cup \Omega_- \cup \Omega_0$, where $\Omega_{\pm}$ and $\Omega_0$
are the subsets of configurations with magnetization $\lim_{V
\rightarrow \infty} \frac{1}{V}\int_V d \vec{x}\phi(\vec x)$
positive, negative and vanishing, respectively. Denoting by
$\rho_{\pm}[\phi(\vec x)]$ the two broken symmetry pure states,
all equilibrium states are the convex linear combinations of
$\rho_{\pm}$. In particular, the Gibbs state is the symmetric
mixture $\rho_G[\phi(\vec x)]=\frac{1}{Z}\exp (-{\cal H}[\phi(\vec
x)]/T)= \frac{1}{2}\rho_+[\phi(\vec x)] +
\frac{1}{2}\rho_-[\phi(\vec x)]$. The $\Omega_{\pm}$ components
are the domains of attraction of the pure states with
$\rho_+(\Omega_+)= \rho_-(\Omega_-)= 1$ and $\Omega_0$ is the
border in between them, with zero measure in any of the
equilibrium states.

When ergodicity is broken, quite different behaviors may
arise~\cite{Palmer}  depending on the initial condition
$\rho_0[\phi(\vec x)]= \rho([\phi(\vec x)],t=0)$. Here, we
consider the three cases relevant for what follows, assuming that
there are not explicit symmetry breaking terms in the equation of
motion:

\begin{enumerate}

\item {\it equilibration to a pure state}

if $\rho_0(\Omega_{+})=1$ or $\rho_0(\Omega_{-})=1$, in the time evolution
configurations are sampled from either one of $\Omega_{\pm}$ and
$\rho([\phi(\vec x)],t)$ equilibrates to the time independent pure
state $\rho_{\pm}[\phi(\vec x)]$ within the finite relaxation time
$t_{eq} \sim \xi^z$, where $\xi$ is the equilibrium correlation
lenght and $z$ is the dynamic exponent entering the domain growth
law, to be defined shortly. The correlation function is the same
in the two ergodic components and, after equilibration, is time
translation invariant

\be
   C_{st}(\vec{r},t-s)=\langle\phi(\vec x,t)\phi(\vec x',s)\rangle_{\pm}-M^2\label{1.4}
\ee

where $\langle \phi(\vec x)\rangle _{\pm} = \pm M$ is the
spontaneous magnetization. For large distances $r \gg \xi$ and
time separations $t-s \gg t_{eq}$, the clustering property
$\langle \phi(\vec x,t)\phi(\vec x',s)\rangle _{\pm}$ $=\langle
\phi(\vec x,t) \rangle _{\pm} \langle\phi(\vec x',s)\rangle
_{\pm}$ is obeyed and the correlations decay to zero, as required
by ergodicity.

\item {\it equilibration to the Gibbs state}

if $\rho_0(\Omega_+)=\rho_0(\Omega_-) =1/2$, then configurations
are sampled evenly from both disjoint components  $\Omega_+$ and
$\Omega_-$. The probability density $\rho([\phi(\vec x)],t)$
equilibrates now to the Gibbs state $\rho_G[\phi(\vec x)]$ with
the same relaxation time $t_{eq}$ as in the relaxation to the pure
states. Broken ergodicity shows up in the large distance and in
the large time properties of the correlation function. After
equilibration, one has

\be
   C_G(\vec{r},t-s)=C_{st}(\vec{r},t-s)+M^2\label{1.9}
\ee

from which follows that correlations do not vanish asymptotically
or that the clustering property is not obeyed

\be
   \lim_{r\rightarrow\infty}C_G(\vec{r},t-s)=\lim_{(t-s)\rightarrow\infty}C_G(\vec{r},t-s)=M^2.\label{1.10}
\ee

\item {\it falling out of equilibrium over the border~\cite{Laloux,Newman}}

if $\rho_0(\Omega_0)=1$, for the infinite system $\rho(\Omega_0,t)
=1$ also at any finite time after the quench. Namely, the system
does not equilibrate since in any equilibrium state the measure of
$\Omega_0$ vanishes. Phase ordering corresponds to this case. In
fact, the system is initially prepared in equilibrium at very high
temperature (for simplicity $T_I=\infty$) and at the time $t=0$ is
suddenly quenched to a final temperature $T$ below $T_C$. In the
initial state the probability measure over phase space is uniform
$\rho_0 [\phi(\vec x)] = 1/ | \Omega |$, implying that the initial
configuration at $t=0$ belongs almost certainly to $\Omega_0$,
since with a flat measure $| \Omega_0 |$ is overwhelmingly larger
than $| \Omega_{\pm}|$.

The morphology of typical configurations visited as the system
moves over $\Omega_0$ is a patchwork of domains of the two
competing equilibrium phases, which coarsen as the time goes on.
The typical size of domains grows with the power law $L(t) \sim
t^{1/z}$, where $z=2$ (independent of dimensionality) for dynamics
with non conserved order parameter~\cite{Bray94}, as it will be
consider here. The sampling of configurations of this type is
responsible of the peculiar features of phase ordering. At a given
time $t_w$ there remains defined a length $L(t_w)$ such that for
space separations $r \ll L(t_w)$ or for time separations $t-t_w
\ll t_w$ intra-domains properties are probed. Then, everything
goes as in the case 2 of the equilibration to the Gibbs state,
ergodicity looks broken and the correlation function obeys
Eq.~(\ref{1.9}). Conversely, for  $r \gg L(t_w)$ or  $t/t_w \gg
1$, inter-domains properties are probed, ergodicity is restored
(as it should be, since evolution takes place within the single
ergodic component  $\Omega_0$) and eventually the correlation
function decays to zero. However, the peculiarity is that if the
limit $t_w \rightarrow \infty$ is taken before $r \rightarrow
\infty$, in the space sector ergodicity remains broken giving
rise, for instance, to the growth of the Bragg peak in the equal
time structure factor.

According to this picture, the correlation function can be written
as the sum of two contributions

\be
   C(\vec{r},t,s)=C_{st}(\vec{r},t-s)+C_{ag}(\vec{r},t,s)\label{1.131}
\ee

where the first one is the stationary contribution of
Eq.~(\ref{1.4}) describing equilibrium fluctuations in the pure
states and the second one contains all the out of equilibrium
information. The latter one is the correlation function of
interest in the theory of phase ordering where, in order to
isolate it, zero temperature quenches are usually considered as a
device to eliminate the stationary component. It is now well
established that $C_{ag}(\vec{r},t,s)$ obeys scaling in the
form~\cite{Furukawa}

\be
   C_{ag}(\vec{r},t,s)=\widehat{C}(r/L(s),t/s)\label{1.132}
\ee

with $\widehat{C}(x,y)=M^2$ for $x<1$ and $y \sim 1$, while

\be
   \widehat{C}(r/L(s),t/s)\sim(t/s)^{-\lambda/z}h(r/L(s))
\ee

for large time separation~\cite{Bray94}, where $\lambda$ is the
Fisher--Huse exponent.

\end{enumerate}

\section{Zero field cooled magnetization}

Let us next consider what happens when a time independent external
field $h(\vec{x})$ is switched on at the time $t_w$. To linear
order the expectation value of the order parameter at the time $t$
is given by

\be
   \langle\phi(\vec{x},t)\rangle_{h}=\langle\phi(\vec{x},t)\rangle_{0}+\int d\vec{x'}\chi(\vec{x}-\vec{x'},t,t_w)h(\vec{x'})\label{2.2}
\ee

\noindent and if $h(\vec{x})$ is random with expecations
$\overline{h(\vec{x})}=0$, $\overline{h(\vec{x})h(\vec{x'})}=
h_0^2\delta(\vec{x}-\vec{x'})$ then one has

\be
   \chi(\vec{x}-\vec{y},t,t_w)=h_0^{-2}\overline{\langle\phi(\vec{x},t)\rangle_{h}h(\vec{y})}.\label{2.3}
\ee

\indent Namely, ZFC is the correlation at the time $t$ of the
order parameter with the random external field.

Going to the three processes considered above

\begin{enumerate}
\item after equilibration in the pure state has occurred and the
stationary regime has been entered, the order parameter correlates
with the external field via the equilibrium thermal fluctuations.
The fluctuation dissipation theorem (FDT) is obeyed

\be
   \chi_{st}(\vec{r},t-t_w)=\frac{1}{T}\left[C_{st}(\vec{r},t-s=0)-C_{st}(\vec{r},t-t_w)\right]\label{2.4}
\ee

and since $C_{st}(\vec{r},t-t_w)$ decays to zero for $t-t_w >
t_{eq}$, over the same time scale $\chi_{st}(\vec{r},t-t_w)$
saturates to

\be
   \lim_{t\rightarrow\infty}\chi_{st}(\vec{r},t-t_w)=\chi_{eq}(\vec{r})=\frac{1}{T}C_{eq}(\vec{r})\label{2.5}
\ee

which is the susceptibility computed in the final equilibrium
state $\rho_{\pm}[\phi(\vec{x})]$.

\item As far as ZFC is concerned, there is no difference between
the relaxation to the mixed Gibbs state and the relaxation to a
pure state. Hence, FDT is satisfied and can be written both in
terms of $C_{st}$ or $C_G$ since, as Eq.~(\ref{1.9}) shows, they
differ by a constant.

\item In the phase ordering process the system stays out of
equilibrium, so it useful to write ZFC as the sum of two
contributions~\cite{Bouchaud97}

\be
   \chi(\vec{r},t,t_w)=\chi_{st}(\vec{r},t-t_w)+\chi_{ag}(\vec{r},t,t_w)\label{2.7}
\ee

where $\chi_{st}(\vec{r},t-t_w)$ satisfies  Eq.~(\ref{2.4}) and
$\chi_{ag}(\vec{r},t,t_w)$ represents the additional out of
equilibrium response. In connection with this latter contribution
there are two basic questions

i) how does it behave with time

ii) what is the relation between  $\chi_{ag}$ and  $C_{ag}$, if any.

\end{enumerate}

\subsection{Scaling hypothesis}

Since ZFC measures the growth of correlation between the order
parameter and the external field, the first question raised above
addresses the problem of an out of equilibrium mechanism for this
correlation, in addition to the thermal fluctuations accounting
for $\chi_{st}$. Restricting attention from now on, for
simplicity, to the case of coincident points ($\vec{r} = 0$) and
dropping therefore the space dependence, the starting point for
the answer is the assumption of a scaling form

\be
   \chi_{ag}(t,t_w)\sim t_w^{-a_{\chi}}\widehat{\chi}_{ag}(t/t_w)\label{2.800}
\ee

\noindent which is the counterpart of Eq.~(\ref{1.132}) for the
correlation function.

\indent The next step is to make statements on the exponent
$a_{\chi}$ and on the scaling function $\widehat{\chi}_{ag}(x)$  .
There exists in the literature an estimate of $a_{\chi}$ based on
simple reasoning. What makes phase ordering different from
relaxation in the pure or in the Gibbs state is the existence of
defects. The simplest assumption is that $\chi_{ag}(t,t_w)$ is
proportional to the density of
defects~\cite{Barrat98,Franz98,Ricci99}. This implies

\be
   a_{\chi}=\delta\label{2.82}
\ee

\noindent where the exponent $\delta$ regulates the time
dependence of the density of defects $\rho_{defect}(t)  \sim
L(t)^{-n} \sim t^{-\delta}$, namely

\be
   \delta=n/z\label{3.2}
\ee

\noindent with $n=1$ for scalar and $n=2$ for vector order
parameter~\cite{Bray94}.

According to this argument $a_{\chi}$ should be independent of
dimensionality. This conclusion is not corroborated by the
available exact, approximate and numerical results. On the basis
of exact analytical solutions for the $d=1$ Ising
model~\cite{Lippiello2000,Godreche} and for the large $N$
model~\cite{Corberi2002}, approximate analytical results based on
the gaussian auxiliary field (GAF)
approximation~\cite{Berthier99,Corberi2001} and numerical results
from simulations of the Ising
model~\cite{Corberi2001,prl,preprint} with $d=2,3,4$, the
following general formula for $a_{\chi}$ has been obtained

\be\label{3.8}
    a_{\chi} = \left \{ \begin{array}{ll}
        \delta \left( \frac{d-d_L}{d_U-d_L} \right ) \qquad $for$ \qquad d < d_U \\
        \delta  \qquad $with log corrections for$ \qquad d = d_U \\
        \delta  \qquad $for$ \qquad  d > d_U
        \end{array}
        \right .
        \label{3.1}
\ee

\noindent where $d_L$ and $d_U > d_L$ do depend on the system in
the following way

\begin{itemize}

\item $d_L$ is the dimensionality where  $a_{\chi}=0$.
In the Ising model $d_L=1$, while in the large
$N$ model $d_L =2$. The speculation is that in general $d_L=1$ for
systems with discrete symmetry and $d_L =2$ for systems with
continous symmetry, therefore suggesting that $d_L$ coincides with the lower
critical dimensionality of equilibrium critical phenomena,
although the reasons for this identification are far from clear.

\item $d_U$ is a value of the dimensionality specific of ZFC and
separating $d <d_U$, where $a_{\chi}$ depends on $d$, from $d >d_U$
where $a_{\chi}$ is independent of dimensionality and
Eq.~(\ref{2.82}) holds. The existence of $d_U$ is
due~\cite{preprint} to a mechanism, i.e. the existence of a
dangerous irrelevant variable, quite similar (including
logarithmic corrections) to the one leading to the breaking of
hyperscaling above the upper critical dimensionality in static
critical phenomena. However, $d_U$ cannot be identified with the
upper critical dimensionality since we have found, so far,  $d_U
=3$ in the Ising model and  $d_U =4$ in the large $N$ model.
In the scalar case it may be argued~\cite{generic} that $d_U$ coincides with the
dimensionality $d_R=3$ such that interfaces do roughen for
$d \leq d_R$ and do not for $d > d_R$. A
general criterion for establishing the value of $d_U$ is not yet
known.

\end{itemize}

\begin{figure}[htpb]
\begin{center}
   \includegraphics[width=8cm]{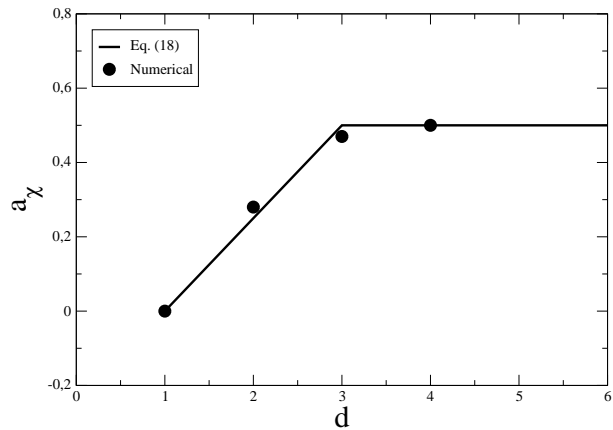}
   \caption{Exponent $a_{\chi}$ in the Ising model at various
dimensionalities. The continous line represents Eq.~(\ref{3.8}),
while the dots are the values from the exact
solution\cite{Lippiello2000} of the model at $d=1$ and from
simulations\cite{preprint} at $d=2,3,4$.} \label{Fig1}
\end{center}
\end{figure}

The validity of Eq.~(\ref{2.800}) with $a_{\chi}$ given by
Eq.~(\ref{3.1}) has been checked, in addition to the cases where
analytical results are available, with very good accuracy in the
simulations of the Ising model~\cite{preprint}. The values of
$\delta$, $d_L$ and $d_U$ obtained for the different systems are
collected in Table \ref{tabella} and the behavior of $a_{\chi}$ as
dimensionality is varied is displayed in Fig.\ref{Fig1}.

\begin{table}[htbp]
\begin{tabular}{||l||c|c|c||}
    \hline
    \hline           &  Ising &  GAF &  $N=\infty$ \\
    \hline  $\delta$ &   1/2  &  1/2 & 1 \\
    \hline  $d_L$    &   1    &  1   &  2 \\
    \hline  $d_\chi$ &  3    & 2   & 4 \\
    \hline
    \hline
\end{tabular}\caption{Parameters entering Eq.(\ref{3.8}) in various models.}
\label{tabella}
\end{table}

At this point, a comment is in order. From Eqs.~(\ref{2.7})
and~(\ref{2.800}) follows that the existence of out of equilibrium
degrees of freedom, or defects, generates the aging contribution
to ZFC. However, when  $a_{\chi} >0$ this is bound to disappear as
time becomes large, either $t_w \rightarrow \infty$ or  $t/t_w
\rightarrow \infty$. The vanishing of this contribution is fastest
for $d > d_U$, where $a_{\chi} = \delta$, and becomes slower and
slower below $d_U$, as $d_L$ is approached. Nonetheless, as long
as  $a_{\chi} >0$, eventually $\chi_{ag}$ disappears and
Eq.~(\ref{2.5}) holds also for phase ordering processes for all $d
> d_L$. This is no longer true for $d \leq d_L$ where $a_{\chi}
\leq 0$ and, consequently, $\chi_{ag}$ gives a contribution to the
response which persists also in the asymptotic time region leading
to

\be
   \lim_{t\rightarrow\infty}\chi(t,t_w)>\chi_{eq}\label{3.100}
\ee

\noindent where $\chi_{eq}$ stands for $\chi_{eq}(\vec{r}=0)$. In
the following we will consider $d \geq d_L$.

\subsection{Fluctuation dissipation relation}

Let us now come to the second question. A very useful tool for the
study of slow relaxation phenomena has been introduced by
Cugliandolo and Kurchan~\cite{Cugliandolo93} through the off
equilibrium fluctuation dissipation relation. This can be
introduced in the following way. Given that $C(t,t_w)$ is a
monotonously decreasing function of $t$, for fixed $t_w$ it is
possible to invert it and write

\be
   \chi(t,t_w)=\widetilde{\chi}(C(t,t_w),t_w).\label{1.10}
\ee

\indent Then, if for a fixed value of $C(t,t_w)$ there exists the
limit

\be
   \lim_{t_w\rightarrow\infty}\widetilde{\chi}(C,t_w)=S(C)\label{1.20}
\ee

\noindent the function $S(C)$ gives the fluctuation dissipation
relation. In the particular case of equilibrium dynamics, FDT is
recovered and $S(C)=[C(0)-C]/T$. Originally introduced in the
study of the low temperature phase of spin glass mean-field
models, the fluctuation dissipation relation has been found in
many other instances of slow relaxation~\cite{Crisanti2002}.

Now, in order to search for $S(C)$ in the case of phase ordering,
let us set $\vec{r}=0$ in Eq.~(\ref{1.132}) and let us eliminate
$t/t_w$ between $\widehat{\chi}_{ag}$ and $C_{ag}$ obtaining

\be
   \chi_{ag}(t,t_w)\sim t_w^{-a_{\chi}}\widetilde{\chi}_{ag}(C_{ag}).\label{2.8}
\ee

Then, from Eqs. ~(\ref{2.7},\ref{2.4},\ref{2.8}) one can write the
general relation

\be
   \chi(t,t_w)=\frac{1}{T}\left[C_{st}(0)-C_{st}(t-t_w)\right]+t_w^{-a_{\chi}}\widetilde{\chi}_{ag}(C_{ag}).\label{2.9}
\ee

Using the identity $\left [ C_{st}(0) - C_{st}(t-t_w)\right ] =
\left [ C_{st}(0)+ M^2  - C_{st}(t-t_w) - M^2 \right ]$ and
considering that in the time interval where $C_{st}(t-t_w) \neq
0$, i.e. for short times, one can replace $C_{ag}(t/t_w)$ with
$M^2$ or equivalently $ C_{st}(t-t_w) + M^2 = C(t,t_w)$, the above
equation can be rewritten as

\be
   \chi(t,t_w)=\widetilde{\chi}_{st}(C)+t_w^{-a_{\chi}}\widetilde{\chi}_{ag}(C_{ag})\label{2.9}
\ee

where the function $\widetilde{\chi}_{st}(C)$ is defined by

\be
    T\widetilde{\chi}_{st}(C) = \left \{ \begin{array}{ll}
        \left [ C(0) - C(t,t_w) \right ]  \qquad $for$ \qquad  M^2 \leq C \leq C(0) \\
        \left [ C(0) - M^2 \right ]  \qquad $for$ \qquad C < M^2.
        \end{array}
        \right .
        \label{2.10}
\ee

Therefore, from Eq.~(\ref{2.9}) we have, first of all, that for
phase ordering systems the fluctuation dissipation relation exists
only if $a_{\chi} \geq 0$ (i.e. for $d \geq d_L$) and that for
$a_{\chi} > 0$

\be
   S(C)=\widetilde{\chi}_{st}(C).\label{1.30}
\ee

For $a_{\chi} = 0$ a little more care is needed.
Equation~(\ref{2.9}) yields $\chi(t,t_w) =
\widetilde{\chi}_{st}(C) + \widetilde{\chi}_{ag}(C_{ag})$.
Recalling that $a_{\chi} = 0$ occurs at  $d = d_L$, which
coincides with the lower critical dimensionality, in order to have
a phase ordering process a quench to $T=0$ is required. This, in
turn, implies $C_{st}(t,s) =0$ and $C_{ag}(t,s) =C(t,s)$.
Therefore, using Eq.~(\ref{2.10}) we have

\be
   S(C)=\chi_{eq}^*+\widetilde{\chi}_{ag}(C)\label{1.40}
\ee

\noindent where $\chi_{eq}^* = \lim_{T \rightarrow 0}[C(0) -
M^2]/T$ is the $T=0$ equilibrium susceptibility, which vanishes
for hard spins while is different from zero for soft spins.
Physical implications of these results are discussed in the next
section.

\section{Statics from dynamics}

One of the main reasons of interest in the fluctuation dissipation
relation is that it may provide a link between static and dynamic
properties. This was first found by Cugliandolo and
Kurchan~\cite{Cugliandolo93} for mean-field spin glasses and then
established in general by Franz {\it et al.}~\cite{Franz98} for
slowly relaxing systems.

Let us first introduce the overlap probability function
$\widetilde{P}(q)$ in the equilibrium state obtained when the
perturbation responsible of $\chi(t,t_w)$ is switched off. The
question is how is $\widetilde{P}(q)$ related to the unperturbed
overlap function $P(q)$. If $P(q) =\widetilde{P}(q)$ the system is
stochastically stable~\cite{Guerra}. A milder statement of
stochastic stability is that $\widetilde{P}(q)$ coincides with
$P(q)$ up to the effects of a global symmetry which might be
removed by the perturbation. In  the Ising case, where the
perturbation breaks the up-down symmetry, defining \be
\widehat{P}(q)= 2\theta(q)P(q) \label{4.00} \ee the system is
stochastically stable in the sense that $\widetilde{P}(q) =
\widehat{P}(q)$. If the system is not stochastically stable,
$\widetilde{P}(q)$ is not related neither to $P(q)$ nor to
$\widehat{P}(q)$. As we shall see, this is the case of the mean
spherical model.

The statement is that

\begin{enumerate}

\item if $S(C)$ exists

\item if $\lim_{t \rightarrow \infty} \chi(t,t_w) = \chi_{eq}$

\item if the system is stochastically stable

\end{enumerate}

then one has

\be
   \left.-T\frac{d^2 S(C)}{d C^2}\right|_{C=q}=\widetilde{P}(q).\label{4.3}
\ee

We may now check if this chain of connections applies to phase
ordering. In replica symmetric low temperature states, as for
instance in ferromagnetic systems,  the overlap function is always
trivial and we have

\be
   P(q) = \frac{1}{2} \left [ \delta (q-M^2) + \delta (q+M^2)
\right ] \label{4.4}
\ee

with
\begin{equation}
\widetilde{P}(q) = \widehat{P}(q) = \delta (q-M^2).
\label{4.4bis}
\end{equation}
Computing the derivative in the left hand side of Eq.~(\ref{4.3})
and using Eqs.~(\ref{1.30}) and~(\ref{1.40}), for $d> d_L$ we find

\be
   \left. -T \frac{d^2 S(C)}{d C^2} \right |_{C=q} = \delta
(q-M^2). \label{4.6}
\ee
Therefore, Eq.~(\ref{4.3}) is satisfied.

\subsection{Failure at $d=d_L$}

Let us, next, go to the case $d= d_L$. For the sake of
definiteness, we consider the case of the Ising model with $d=1$. In
order to make compatible the two requirements of having an ordered
equilibrium state and a well defined linear response function,
instead of taking the $T \rightarrow 0$ limit it is necessary to
take the limit of an infinite ferromagnetic
coupling~\cite{Lippiello2000}. Then, $P(q)$ and $\widetilde{P}(q)$
are given by Eqs.~(\ref{4.4}) and~(\ref{4.4bis}) with $M^2=1$ at
all temperatures. On the other hand, for any $T$ we also
have~\cite{Lippiello2000}

\begin{equation}
T\widetilde{\chi}_{ag}(C) = \frac{\sqrt{2}}{\pi} \arctan \left [ \sqrt{2}\cot \left (
\frac{\pi}{2} C \right ) \right ].
\label{4.5}
\end{equation}

\noindent This gives

\be
   \left.-T\frac{d^2 S(C)}{d C^2} \right |_{C=q} = \frac{\pi
\cos (\pi q/2)\sin (\pi q/2)}{[2 -\sin (\pi q/2)]^2}.
\label{4.6bis}
\ee

\noindent Hence, it is clear that Eq.~(\ref{4.3}) is not verified.
The reason is that the second of the above conditions required for
establishing the connection is not satisfied. In fact, from
Eqs.~(\ref{2.9}) and~(\ref{4.5}), keeping in mind that the limits
$t \rightarrow \infty$ and $C \rightarrow 0$ are equivalent, we
have

\be
   \lim_{t\rightarrow\infty}T\chi(t,t_w)=1/\sqrt{2}\label{4.8}
\ee

\noindent which is responsible of Eq.~(\ref{3.100}) and,
therefore, of the violation of condition (2) above
Eq.~(\ref{4.3}), since in this case $\chi_{eq}=0$.

\subsection{Failure by stochastic instability}

\begin{figure}[htpb]
\begin{center}
   \includegraphics[width=8cm]{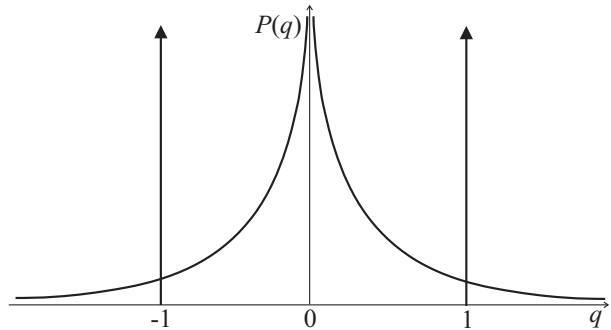}
   \caption{Overlap distribution for mean spherical model with
$M^2=1$. The arrows represent the $\delta$ functions of the
overlap distribution for the spherical model.}\label{Fig2}
\end{center}
\end{figure}

An interesting example~\cite{Fusco}, where statics cannot be reconstructed from dynamics because the
third requirement of stochastic
stability is not satisfied, comes from the spherical model. More precisely,
one must consider in parallel the original version of the spherical model (SM) of Berlin and
Kac~\cite{Berlin} and the mean spherical model (MSM) introduced by Lewis and Wannier~\cite{Lewis}, with the
spherical constraint treated in the mean. These two models are equivalent above
but not below  $T_C$~\cite{Kac}. The low temperature states are quite different, with a bimodal
order parameter probability distribution in the SM case and a gaussian distribution
centered in the origin in the MSM case.  The corresponding  overlap functions are also very
different~\cite{Fusco}. Considering, for simplicity, $T=0$ one has
\be
    P(q) = \left \{ \begin{array}{ll}
        \frac{1}{2} \left [ \delta (q-M^2) + \delta (q+M^2) \right ]
\qquad $for SM$ \\
        \frac{1}{\pi M^2}K_0(|q|/ M^2) \qquad $for MSM$
        \end{array}
        \right .
        \label{5.1}
        \ee
where $K_0$ is a Bessel function of imaginary argument
(Fig.\ref{Fig2}).

However, after switching off an external field, one finds for {\it
both} models $\widetilde{P}(q) = \delta (q-M^2)$. This means that
stochastic stability holds for SM but not for MSM.

On the other hand, the relaxation properties are the same in the two models, both above and below
$T_C$ if the thermodynamic limit is taken before the $t \rightarrow \infty$ limit~\cite{Fusco}.
Then, the linear
response function is the same for both models and obeys Eq.~(\ref{2.9}) with
$a_{\chi}$ given by  Eq.~(\ref{3.1}), where $\delta$, $d_L$ and  $d_U$ are the same as for
the large $N$ model (Table I). Hence, we have that although Eq.~(\ref{4.3}) is satisfied for
both models, nonetheless statics and dynamics are connected only in the SM case,
where $\widetilde{P}(q) = \widehat{P}(q)$. Instead, this is not possible in the MSM case
where $\widetilde{P}(q) \neq \widehat{P}(q)$.

\section{Conclusions}

We have shown that although the low temperature equilibrium state in phase ordering
systems is simple, ZFC in quenches below $T_C$ displays non trivial features.
Summarising, these are

\begin{itemize}

\item The aging component $\chi_{ag}(t,t_w)$ obeys the scaling form~(\ref{2.800})
with an exponent $a_{\chi}$ dependent on dimensionality according to
Eq.~(\ref{3.1}).

\item The fluctuation dissipation relation $S(C)$ is trivial, in the sense that it is
consistent with a replica symmetric equilibrium state, for $d > d_L$. However, as
$d \rightarrow d_L$ from above the non trivial contribution due to $\chi_{ag}(t,t_w)$
persists for longer and longer times as $a_{\chi} \rightarrow  0$.

\item For $d = d_L$ the fluctuation dissipation relation is non trivial, since the
aging contribution does not disappear asymptotically. From this follows i) that
there is no connection between static and dynamic properties and ii) that on the
basis of the behavior of ZFC is not possible to establish that the equilibrium
state is replica symmetric.

\item The mean spherical model provides an explicit example of a system whose low
temperature equilibrium state is not stochastically stable, producing the failure of
the link between static and dynamic properties for all dimensionalities.

\item Extensive numerical simulations for ferromagnetic systems with scalar, vector,
conserved and non conserved order parameter, at various space dimensionalities are under way.
Preliminary results do show that the behavior of the response function above
illustrated might well be generic for phase ordering systems~\cite{generic}.

\end{itemize}

\section*{Acknowledgements}

This work has been partially supported from MURST through
PRIN-2002.

\end{document}